\documentclass[twocolumn,showpacs,preprintnumbers,amsmath,amssymb,superscriptaddress,prd]{revtex4}

\usepackage{graphicx}
\usepackage{dcolumn}
\usepackage{bm}
\usepackage{amsmath,amssymb}

\begin{document}

\preprint{APS/123-QED}

\title{Constraining Extended Gravity Models by S2 star orbits around the Galactic Centre}

\author{S. Capozziello}
\email[Corresponding author:]{capozziello@na.infn.it}
\affiliation{Dipartimento di  Fisica, Universit\`{a} di Napoli "Federico II",
 Compl. Univ. di Monte S. Angelo, Edificio G, Via Cinthia, I-80126, Napoli, Italy,}
\affiliation{Istituto Nazionale di  Fisica Nucleare (INFN) Sez. di Napoli, Compl. Univ. di Monte S. Angelo,
Edificio G, Via Cinthia, I-80126, Napoli, Italy,} \affiliation{Gran
Sasso Science Institute (INFN),  Viale F. Crispi, 7, I-67100,
L'Aquila, Italy.}

\author{D. Borka}
\affiliation{Atomic
Physics Laboratory (040), Vin\v{c}a Institute of Nuclear Sciences,
University of Belgrade, P.O. Box 522, 11001 Belgrade, Serbia.}

\author{P. Jovanovi\'{c}}
\affiliation{Astronomical Observatory, Volgina 7, 11060 Belgrade,
Serbia.}

\author{V. Borka Jovanovi\'{c}}
\affiliation{Atomic Physics Laboratory (040), Vin\v{c}a Institute of
Nuclear Sciences, University of Belgrade, P.O. Box 522, 11001
Belgrade, Serbia.}

\date{\today}

\begin{abstract}
We investigate the possibility to explain theoretically the observed
deviations of S2 star orbit around the Galactic Centre using
gravitational potentials derived from modified gravity models in
 absence of dark matter. To this aim, an analytic fourth-order
theory of gravity, non-minimally coupled with a massive scalar field
is considered. Specifically, the interaction term is given by
analytic functions $f(R)$ and $f(R,\phi)$ where $R$ is the Ricci
scalar and $\phi$ is a scalar field whose meaning can be related to
further gravitational degrees of freedom. We simulate the orbit of
S2 star around the Galactic Centre in $f(R)$ (Yukawa-like) and
$f(R,\phi)$ (Sanders-like) gravity potentials and compare it with
NTT/VLT observations. Our simulations result in strong constraints
on the range of gravity interaction. In the case of analytic
functions $f(R)$, we are not able to obtain reliable constraints on
the derivative constants $f_1$ and $f_2$, because the current
observations of S2 star indicated that they may be highly mutually
correlated. In the case of analytic functions $f(R,\phi)$, we are
able to obtain reliable constraints on the derivative constants
$f_0$, $f_R$, $f_{RR}$, $f_{\phi}$, $f_{\phi\phi}$ and $f_{\phi R}$.
The approach we are proposing seems to be sufficiently reliable to
constrain the modified gravity models from stellar orbits around
Galactic Centre.
\end{abstract}

\pacs{04.50.Kd, 04.25.Nx, 04.40.Nr}

\maketitle

\section{Introduction}

Extended Theories of Gravity \cite{capo11} are alternative theories
of gravitational interaction developed from the exact starting
points investigated first by Einstein and Hilbert and aimed from one
side to extend the positive results of General Relativity  and, on
the other hand, to cure its shortcomings. Besides  other fundamental
issues, like dark energy and quantum gravity, these theories have
been proposed like alternative approaches to Newtonian gravity in
order to explain galactic and extragalactic dynamics without
introducing dark matter \cite{capo12,noji11}.  In particular, the
search for non-Newtonian gravity is part of the quest for
non-Einsteinian physics which consists of searching for deviations
from Special and General Relativity \cite{fisc99,cope04,clif12}.
They are aimed to address conceptual and experimental problems
recently emerged in astrophysics and cosmology from the observations
of the Solar system, binary pulsars, spiral galaxies, clusters of
galaxies and the large-scale structure of the Universe
\cite{capo02,capo03,carr04,leon11,capo10}. In general, these
theories describe gravity as a metric theory with a linear
connection but  there are also affine, or metric-affine formulations
of Extended Theories of Gravity \cite{capo11}. Essentially, they are
based on  straightforward generalizations of the Einstein theory
where the gravitational action (the Hilbert-Einstein action) is
assumed to be linear in the Ricci curvature scalar $R$. $f(R)$
gravity is a type of modified gravity which generalizes Einstein's
General Relativity and it was first proposed in 1970 by Buchdahl
\cite{buch70}. It is actually a family of models, each one defined
by a different function of the Ricci scalar. The simplest case is
just the General Relativity. In the case of $f(R)$ gravity, one
assumes a generic function $f$ of the Ricci scalar $R$ (in
particular, analytic functions) and searches for a theory of gravity
having suitable behavior at small and large scale lengths. As a
consequence of introducing an arbitrary function, there may be
freedom to explain the accelerated expansion and structure formation
of the Universe without adding unknown forms of dark energy or dark
matter. One type of the Extended Theories of Gravity is
characterized by power-law Lagrangians \cite{capo06,capo07}.
Alternative approaches to Newtonian gravity in the framework of the
weak field limit of fourth order gravity theory have been proposed
and constraints on these theories have been discussed
\cite{zakh06,zakh07,nuci07,frig07,bork12,zakh14,bork13,li07}.

Yukawa-like corrections have been obtained in the framework of
$f(R)$  gravity as a general feature of these theories
\cite{stelle,capo09a,card11,iori10}. It is important to stress that
they emerge as exact solutions in the context of Extended Gravity
and are not just put by hand  as phenomenological terms. The
physical meaning of such corrections needs to be confirmed at
different scales: for short distances, Solar system, spiral galaxies
and galaxy clusters. A compilation of experimental, geophysical and
astronomical constraints on Yukawa violations of the gravitational
inverse square law are given in Figs. 9 and 10 from \cite{adel09}
for different ranges. These results show that the Yukawa term is
relatively well constrained for the short ranges. For longer
distances Yukawa corrections have been successfully applied to
clusters of galaxies \cite{capo07b,capo09b,card11}. Lucchesi and
Peron \cite{lucc14} analyzed pericenter general relativistic
precession and gave constraints on exponential potential to Solar
System measurements. However, further tests are needed in order to
set robust constraints on Yukawa corrections. Galactic stellar
dynamics could be of great aid in this program.

S-stars are mainly young early-type stars that closely orbit the
massive compact object at the center of Milky Way, named Sgr
A$^\ast$ \cite{ghez00,scho02,gill09a,gill09b,ghez08,genz10}. These
stars, together with recently discovered dense gas cloud falling
towards the Galactic Centre \cite{gill12}, indicate that the massive
central object is a black hole. In our simulation we will treat
central object like "massive compact object" since our goal was only to study
orbits of stars around Galactic Centre, no matter what is the nature
of the object (black hole or not). For at least one of them, called S2, there
are some observational indications that its orbit maybe deviates from
the Keplerian case due to relativistic precession \cite{gill09a,meye12}.

However, we have to point out that the present astrometric limit is still not sufficient to definitely
confirm such a claim. On the other hand, the astrometric accuracy is
constantly improving from around 10 mas during the first  part of
the observational period, currently reaching less than 1 mas (0.3
mas) see \cite{fritz}.
Furthermore, some  recent studies provide more and more evidence that
the orbit of S2 star is not closing (see e.g. Fig. 2 in \cite{meye12}).
Here, we fitted the NTT/VLT astrometric observations of S2 star,
which  contain  a possible indication for orbital precession around the
massive compact object at Galactic Centre, in order to constrain the
parameters of Sanders-like gravity potential, since this kind of
potential has not been tested at these scales yet. We obtained much
larger orbital precession of S2 star in Sanders-like  gravity than
the corresponding value predicted by General Relativity. In the paper \cite{gill09a}  page 1092,  Fig. 13,
authors presented the Keplerian orbit but they have to move the
position of central point mass to explain orbital precession. In our
 orbit, calculated  by Sanders-like potential for best fitting
parameters, we also obtained precession, but with a fixed position of the
central point mass. In other words, we do not need to move central point mass in order to get the fit.

As a general remark, the orbit of S2 will give astronomers the opportunity to test for various
effects predicted by General Relativity. The orbital precession can
occur due to relativistic effects, resulting in a prograde pericentre
shift or due to a possible extended mass distribution, producing a retrograde
shift \cite{rubi01}. Both prograde relativistic and retrograde
Newtonian pericentre shifts will result in {\it rosette} shaped
orbits \cite{adki07}. We have to stress that the current astrometric
limit is not sufficient to unambiguously confirm such a claim.
Weinberg et al. \cite{wein05} discussed physical experiments
achievable via the monitoring of stellar dynamics near the Galactic
Centre with a diffraction-limited, next-generation, extremely large
telescope (ELT).

The aim of this paper is to give the astronomical constraints on
Extended Theories of Gravity by using the peculiar dynamics of S2
star. In particular, we want to fix the ranges of Yukawa-like
correction parameters adopting the NTT/VLT observations. The paper
is organized as follows. Sec. II is devoted to a short summary of
Extended Gravity in view of the Newtonian limit where Yukawa-like
corrections emerge. The simulated orbits of S2 star in modified
potential  are considered in Sec. III. In particular, we set the
problem of how to constrain the Yukawa-potential parameters. Sec. IV
is devoted to the results of the simulation. Conclusions are drawn
in Sec. V.

\section{Extended Theories of Gravity}

Examples of Extended Theories of Gravity  are higher-order,
scalar-tensor gravities, see, for example
\cite{capo11,noji11,noji07,capo08,capo08b,capo09c,capo10,bisw12,moff05,moff06}.
These theories can be characterized by two main features: $i)$ the
geometry can non-minimally couple to some scalar field; $ii)$
higher-order curvature invariants can appear into the action.
In the first case, we are dealing with scalar-tensor
gravity, and in the second case we have higher-order gravity.
Combinations of non-minimally coupled and higher order terms can
also emerge in effective Lagrangian, producing mixed higher
order/scalar-tensor gravity. A general class of
higher-order-scalar-tensor theories in four dimensions is given by
the effective action \cite{capo11,stab13}:

\begin{eqnarray} \label{V3.1}
 {\cal S}=&&\int d^{4}x\sqrt{-g}\left[f(R,\Box R,\Box^{2}R,\dots,\Box^kR,\phi)+\right.\nonumber\\
 &&\left.\omega(\phi)
\phi_{; \alpha} \phi^{;\alpha}
+\mathcal{X} \mathcal{L}_m\right],
\end{eqnarray}
where $f$ is an unspecified function of curvature invariants and
 scalar field $\phi$ and $\mathcal{X}\,=\,8\pi G$. Here we use the
convention $c\,=\,1$.

The simplest extension of General Relativity  is achieved assuming:

\begin{eqnarray}\label{fr}
R\rightarrow\,f(R)\,,\qquad \omega(\phi)\,=\,0\,
\end{eqnarray}
where the action (\ref{V3.1}) becomes \cite{stab13}:

\begin{equation} \label{s1_fR}
{\cal S}= \int  d^{4}x\sqrt{-g}\left[f(R)+\mathcal{X}\mathcal{L}_m\right]\,.
\end{equation}

A general gravitational potential, with a Yukawa correction, can be
obtained in the Newtonian limit of any analytic $f(R)$-gravity
model. From a phenomenological point of view, this correction
allows to consider as viable this kind of models even at small
distances, provided that the Yukawa correction turns out to be
not relevant in this approximation as in the so called "Chameleon
Mechanism" \cite{capo08b}.

For the sake of simplicity,  one can  assume, however, analytic
Taylor expandable $f(R)$ functions with respect to the value
$R\,=\,0$ that is the Minkowskian background \cite{stab13}:

\begin{eqnarray}\label{sertay}
f(R)\,=\,\sum_{n\,=\,0}^{\infty}\frac{f^{(n)}(0)}{n!}\,R^n\,=\,
f_0+f_1R+\frac{f_2}{2}R^2+...
\end{eqnarray}
It is worth noticing that, at the order ${\mathcal O}(0)$, the field
equations give the condition $f_0 =0$ and then the solutions at
further orders do not depend on this parameter. On the other hand,
considering the first term in $R$, $f_0$ has the meaning of a cosmological constant.

A further step is to analyze the Newtonian limit starting from the
action (\ref{V3.1}) and considering a generic function of Ricci
scalar and scalar field. Then the action becomes \cite{stab13}

\begin{eqnarray}\label{HOGaction}
\mathcal{A}=\int d^{4}x\sqrt{-g}\biggl[f(R,\phi)+\omega(\phi)\,\phi_{;\alpha}\,\phi^{;\alpha}+\mathcal{X}\mathcal{L}_m\biggr]
\end{eqnarray}
 The scalar
field $\phi$ can be  approximated as the Ricci scalar. In particular
we get $\phi\,=\,\phi^{(0)}\,+\,\phi^{(1)}\,+\,\phi^{(2)}\,+\dots$
and the function $f(R,\phi)$ with its partial derivatives ($f_R$,
$f_{RR}$, $f_{\phi}$, $f_{\phi\phi}$ and $f_{\phi R}$) and
$\omega(\phi)$ can be substituted by their corresponding Taylor
series. In the case of $f(R,\phi)$, we have \cite{stab13}:

\begin{eqnarray}
f(R,\phi)\sim f(0,\phi^{(0)})+f_R(0,\phi^{(0)})R^{(1)}+f_\phi(0,\phi^{(0)})\phi^{(1)}..
\end{eqnarray}
and analogous relations for the derivatives are obtained. From the
lowest order of field equations we have \cite{stab13}

\begin{eqnarray}\label{PPN-field-equation-general-theory-fR-O0}
f(0,\,\phi^{(0)})\,=\,0\,,\,\,\,\,\,\,\,\,\,\,f_{\phi}(0,\phi^{(0)})\,=\,0
\end{eqnarray}
and also in this modified fourth order gravity a missing
cosmological component in the action (1) implies that the space-time
is asymptotically Minkowskian (the same outcome as above). Moreover
the ground value of scalar field $\phi$ must be a stationary point
of the potential.

An important remark is due at this point. As discussed in details in
\cite{stab13}, a theory like $f(R,\phi)$ is dynamically equivalent
to $f(R,\Box R)$ and then also the meaning of the scalar field results
clearer being related to the further gravitational degrees of freedom
that come out in Extended Gravity \cite{schmidt1,schmidt2}.

\begin{figure*}[ht!]
\centering
\includegraphics[width=0.45\textwidth]{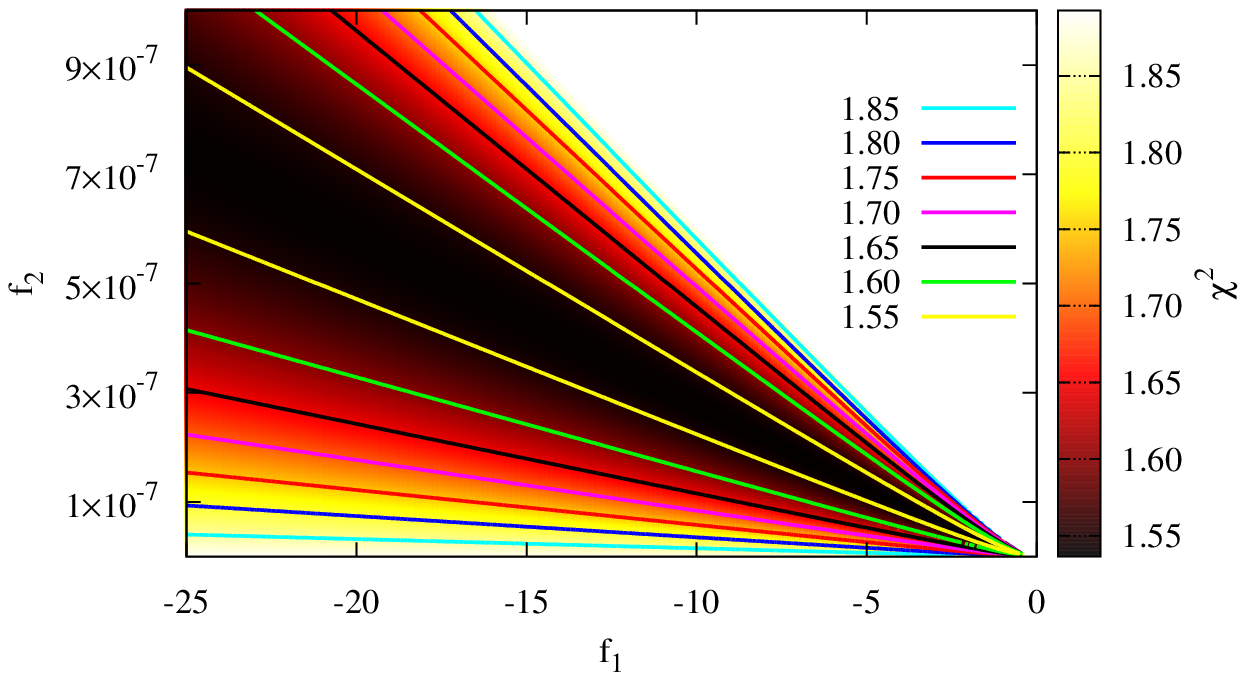}
\hspace{1cm}
\includegraphics[width=0.45\textwidth]{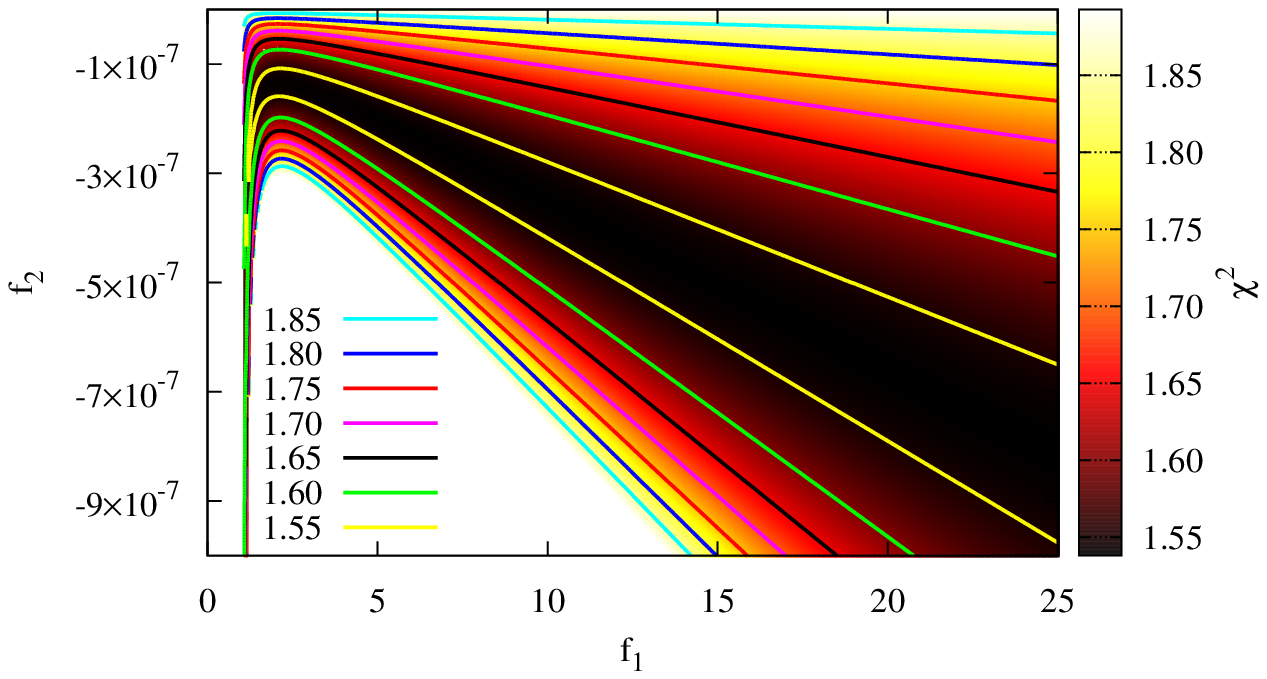}
\caption{(color online) The maps of reduced $\chi^2$ over the $\{f_1
- f_2\}$ parameter space of $f(R)$ gravity in case of NTT/VLT
observations of S2 star which give at least the same
($\chi^{2}=1.89$) or better fits ($\chi^{2}<1.89$) than the
Keplerian orbits. The left panel corresponds to $f_1$ in $[-25, 0]$,
and the right panel to $f_1$ in $[0, 25]$. With decreasing value of
$\chi^{2}$ (better fit) colors in grey scale are darker. A few
contours are presented for specific values of reduced $\chi^2$ given
in figure's legend.}
\label{fig01}
\end{figure*}

\begin{figure*}[ht!]
\centering
\includegraphics[width=0.45\textwidth]{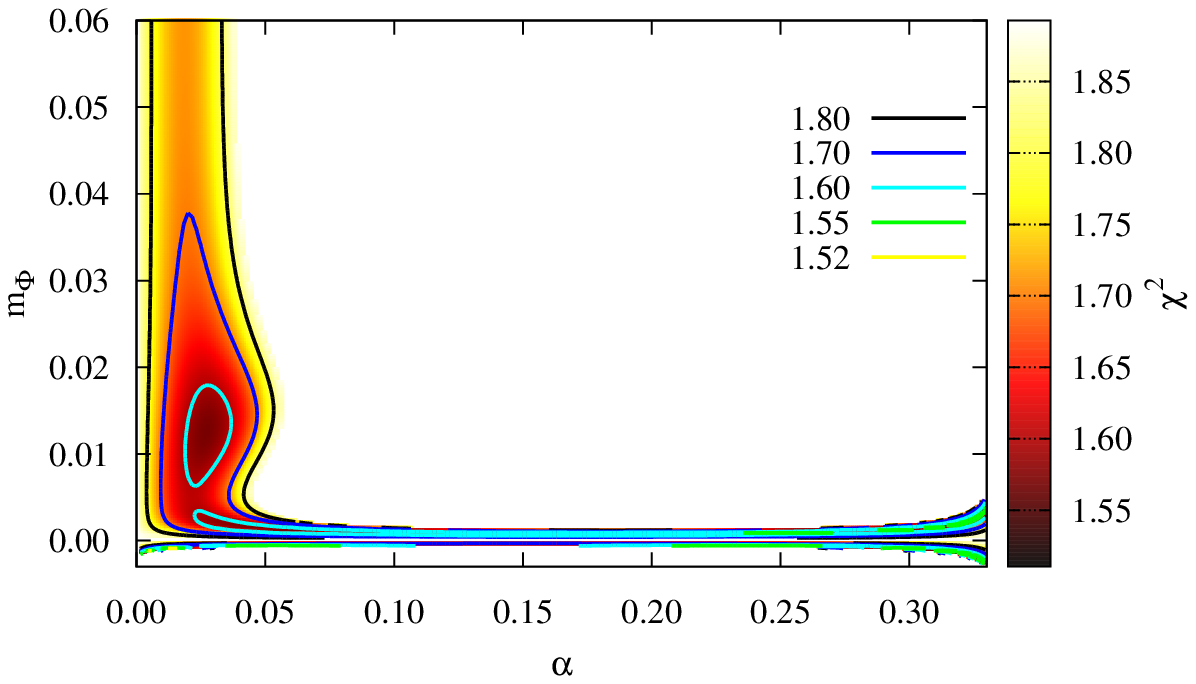}
\hspace{1cm}
\includegraphics[width=0.45\textwidth]{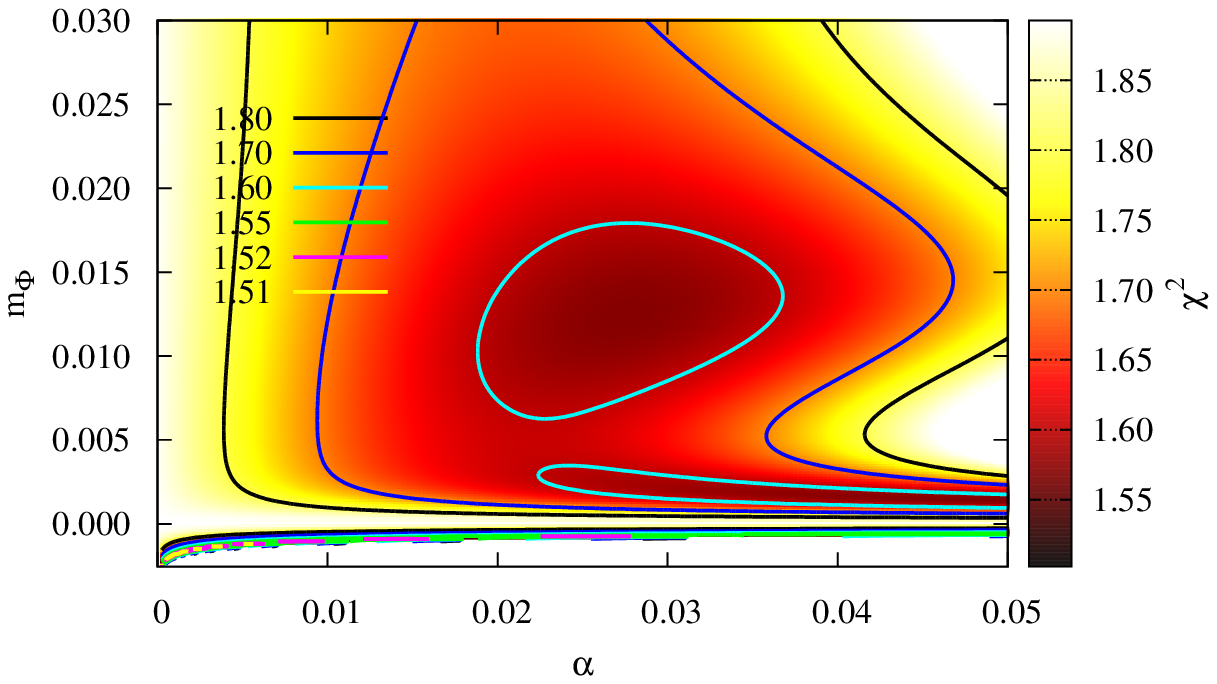}
\caption{(color online) The maps of reduced $\chi^2$ over the
$\{\alpha - m_\phi\}$ parameter space of $f(R,\phi)$ gravity in case
of NTT/VLT observations of S2 star which give at least the same
($\chi^{2}=1.89$) or better fits ($\chi^{2}<1.89$) than the
Keplerian orbits. The left panel corresponds to $m_\phi$ in $[0,
0.06]$ and $\alpha$ in $[0, 0.33]$, and the right panel to the
zoomed range of $m_\phi$ in $[0, 0.03]$ and $\alpha$ in $[0, 0.05]$.
With decreasing value of $\chi^{2}$ (better fit) colors in grey
scale are darker. A few contours are presented for specific values
of reduced $\chi^2$ given in figure's legend.}
\label{fig02}
\end{figure*}

\begin{figure}[ht!]
\includegraphics[width=0.49\textwidth]{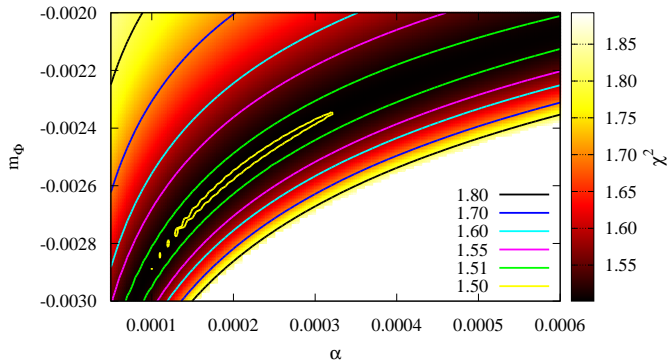}
\caption{(color online) The same as in Fig. 2, but for a narrow
region in the $\{\alpha - m_\phi\}$ parameter space around the
absolute minimum of the reduced $\chi^{2}$. With decreasing value of
$\chi^{2}$ (better fit) colors in grey scale are darker. A few contours
are presented for specific values
of reduced $\chi^2$ given in figure's legend.}
\label{fig03}
\end{figure}

\begin{figure}[ht!]
\includegraphics[width=0.49\textwidth]{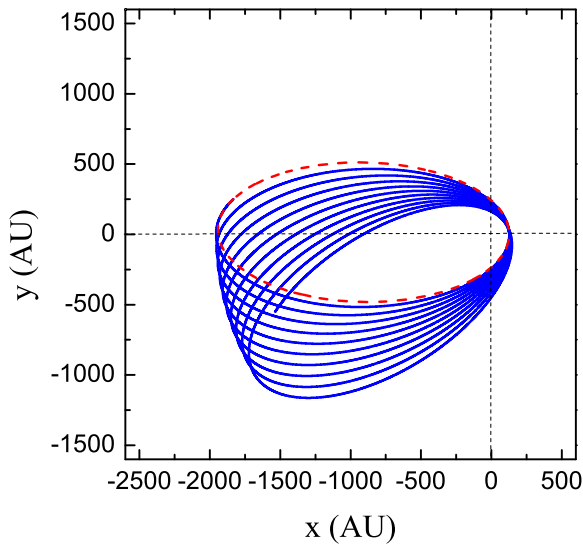}
\caption{(color online) Comparison between the orbit of S2 star in
Newtonian potential (red dashed line) and Sanders-like potential for
the best fit parameters (the absolute minimum of reduced
$\chi^{2}=1.5011$) $\alpha$ = 0.00018 and $m_\phi$ = -0.0026 during
10 orbital periods (blue solid line).}
\label{fig04}
\end{figure}

\begin{figure}[ht!]
\includegraphics[width=0.49\textwidth]{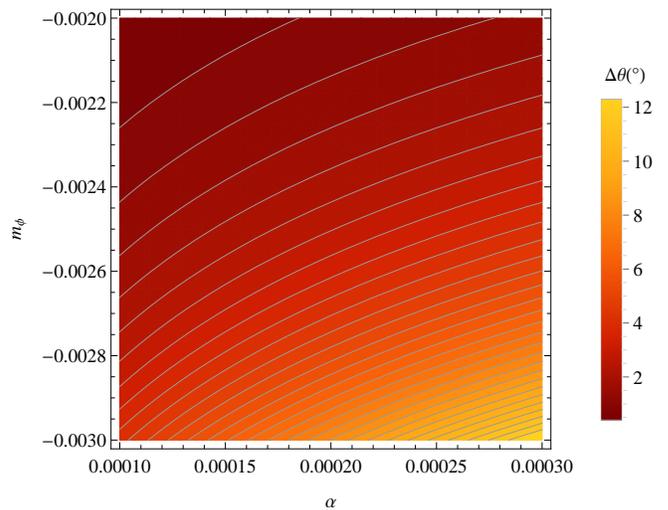}
\caption{(color online) Numerically calculated angle of precession
per orbital period as a function of parameters $\alpha$ in the range
$[0.0001, 0.0003]$ and $m_\phi$ in the range $[-0.003, -0.002]$ in
case of Sanders-like potential. With decreasing value of angle of
precession colors are darker.}
\label{fig05}
\end{figure}

\begin{figure}[ht!]
\includegraphics[width=0.49\textwidth]{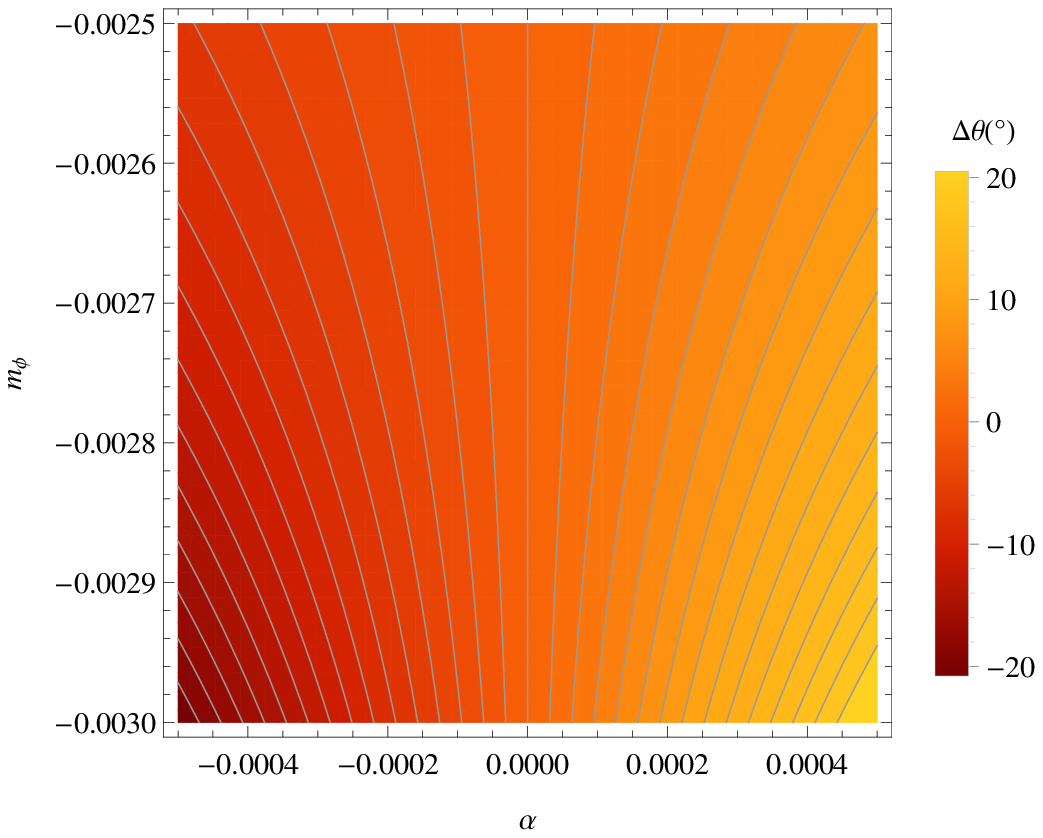}
\caption{(color online) The same as in Fig. \ref{fig05}, but for
$\alpha$ in the range $[-0.0005, 0.0005]$ and $m_\phi$ in the range
$[-0.003, -0.0025]$. The pericenter advance (like in GR) is obtained
for positive $\alpha$, and retrograde precession for negative
$\alpha$. With decreasing value of angle of precession colors are
darker.}
\label{fig06}
\end{figure}

\section{Simulated orbits of S2 star and the Yukawa-like corrections}

In order to constrain the parameters of $f(R)$ and $f(R,\phi)$
models, we simulate orbits of S2 star in Yukawa-like  gravity
potentials and fit them to the astrometric observations obtained by
New Technology Telescope/Very Large Telescope (NTT/VLT) (see Fig. 1
in \cite{gill09a}), which are publicly available as the
supplementary on-line data to the electronic version of the paper
\cite{gill09a} at \cite{url}, and which are aimed to estimate the
distance from the Galactic Centre and to map the inner region of our
Galaxy.

As discussed, in $f(R)$-gravity, the scalar curvature $R$ of the
Hilbert - Einstein action, is replaced by a generic function $f(R)$.
As a result, in the weak field limit \cite{clif05}, the
gravitational potential is found to be Yukawa-like
\cite{capo10,napo12}:

\begin{equation}
\Phi \left( r \right) = -\dfrac{G M}{(1+\delta)r}\left[ {1 + \delta
e^{- \left(\dfrac{r}{\Lambda} \right)}} \right],
\label{equ01}
\end{equation}
where $\Lambda^2=-f_1/f_2$ is an arbitrary parameter (usually
referred to as the range of interaction), depending on the typical scale
of the system under consideration and $\delta=f_1-1$ is a universal
constant.
It is worth noticing that  $\delta$  and $\Lambda$ depend on the parameters of the given $f(R)$ gravity  model. It is important to stress  that a Yukawa-like correction has been invoked several times in the past  \cite{sanders1}.   Such  corrections have been obtained, as a general feature, in the framework of $f(R)$ gravity \cite{capo09a}  and successfully applied to clusters of galaxies setting \cite{capo09b}. In general, one can relate the length-scale $\Lambda$ to the mass of the effective scalar field introduced by the Extended Theory of Gravity and then to the mass and characteristic size of the self-gravitating system \cite{capo11}.
The larger the mass, the smaller will be $\Lambda$ and the faster will be the exponential decay of the correction, i.e. the larger the mass, the quicker the recovery of the Newtonian dynamics. Eq.(\ref{equ01}) then gives us the opportunity to investigate in a unified way the impact of a large class of modified gravity theories, included in which are the Extended Theories, since other details do not have any impact on the galactic scales we are interested in.

In the $f(R,\phi)$-gravity the gravitational potential  is found
by setting the gravitational constant as
\begin{equation}
\label{G}
G\,=\,\left(\frac{2\,\omega(\phi^{(0)})\,\phi^{(0)}-4}{2\,\omega(\phi^{(0)})\,\phi^{(0)}-3}\right)\frac{G_\infty}{\phi^{(0)}}
\end{equation}
where $G_\infty$ is the gravitational constant as measured at
infinity and by imposing
$\alpha^{-1}\,=\,3-2\,\omega(\phi^{(0)})\,\phi^{(0)}$, the gravity
potential is \cite{stab13}:

\begin{eqnarray}
\label{sanders}
\Phi_{ST}(\textbf{x})\,=\,-\frac{G_\infty M}{|\textbf{x}|}\biggl\{1+\alpha\,e^{-\sqrt{1-3\alpha}\,m_\phi |\textbf{x}|}\biggr\}
\end{eqnarray}
and then a Sanders-like potential is fully recovered
\cite{sanders1}. Such a potential has been often used to obtain the
rotation curves of spiral galaxies \cite{sanders2}.
However, it is worth stressing that the  only Sanders potential is unable to
reproduce the rotation curves of spirals without dark matter as pointed out in an accurate study in \cite{dubal}. However, the paradigm remains valid and modifications of Newtonian potential can be investigated in view of addressing the dark matter problem.

We can set
the value of the derivatives of the Taylor expansion as
$f_{R\phi}\,=\,1,\,\,f_{RR}\,=\,0,\,\,f_R\,=\,\phi$ without losing
generality.

The simulated orbits of S2 star in these two potentials can be obtained by
numerical integration of the corresponding differential equations of motion, that is:

\begin{equation}
\mathbf{\dot{r}}=\mathbf{v},\hspace*{0.5cm}
\mu\mathbf{\ddot{r}}=-\triangledown\Phi\left(
\mathbf{r}\right),
\label{2body}
\end{equation}

\noindent where $\mu$ is the reduced mass in the two-body problem.
We assume the following mass and distance for  Sgr A$^\ast$ central
massive compact object around which S2 star is orbiting: $ M   = 4.3
\times10^6 M_\odot$ and $d_\star$ = 8.3 kpc, respectivelly
\cite{gill09a}. Due to simplicity reasons, we perform two-body
simulations and neglect perturbations from other members of the
S-star cluster, as well as from some possibly existing extended
structures composed by visible or dark matter \cite{zakh07}.

We simulate  orbits of S2 star and fit them to the NTT/VLT
astrometric observations for different combinations of {\it a
priori} given values of $f_1$ and $f_2$ in $f(R)$, and $\alpha$ and
$m_\phi$ in $f(R,\phi)$ gravity potentials (below denoted as
$parameter1$ and $parameter2$, respectively). Each simulated orbit
is defined by the following four initial conditions: two components
of initial position and two components of initial velocity in
orbital plane at the epoch of the first observation. For each
combination of $parameter1$ and $parameter2$, we obtain the best fit
initial conditions corresponding to a simulated orbit with the
lowest discrepancy in respect to the observed one. The fitting
itself is performed using LMDIF1 routine from MINPACK-1 Fortran 77
library which solves the nonlinear least squares problems by a
modification of Marquardt-Levenberg algorithm \cite{more80},
according to the following procedure:

\begin{enumerate}
\item In the first iteration we use a guess of initial position $(x_0, y_0)$
and velocity $(\dot{x}_0, \dot{y}_0)$ of S2 star in orbital plane (true orbit)
at the epoch of the first observation;
\item the true positions $(x_i, y_i)$ and velocities $(\dot{x}_i,
\dot{y}_i)$ at all successive observed epochs are then calculated by
numerical integration of equations of motion (\ref{2body}), and projected to
the corresponding positions $(x_i^c, y_i^c)$ in the observed plane
(apparent orbit);
\item discrepancy between the simulated and observed apparent orbit is
estimated by the reduced $\chi^{2}$:
\begin{equation}
    \chi^{2} =
    \dfrac{1}{2N-\nu}{\sum\limits_{i = 1}^N {\left[ {{{\left(
    {\dfrac{x_i^o - x_i^c}{\sigma_{xi}}} \right)}^2}
    + {{\left( \dfrac{y_i^o - y_i^c}{\sigma_{yi}}
    \right)}^2}} \right]} },
\label{chi2}
\end{equation}
where $(x_i^o, y_i^o)$ and $(x_i^c, y_i^c)$ are the corresponding observed and
calculated apparent positions, $N$ is the number of observations, $\nu$ is
number of initial conditions (in our case $\nu=4$), $\sigma_{xi}$ and
$\sigma_{yi}$ are uncertainties of observed positions;
\item the new initial conditions are estimated by the fitting routine and the
steps 2-3 are repeated until the fit is converging, i.e. until the minimum of
reduced $\chi^{2}$ is achieved.
\end{enumerate}
Finally, we kept the values of $parameter1$ and $parameter2$ for which
the smallest value of minimized reduced $\chi^{2}$ is obtained, in other
words, which results with the best fit simulated orbit of S2 star with the
lowest discrepancy with respect to the observed one.

\section{Results and Discussion}

Our point is that the Yukawa-like correction, coming from  $f(R)$
gravity, can  be used in order to fix the coefficients in the
expansion (\ref{sertay}). For the expansion up to the second order,
we have 2 parameters to fix. Orbits of S2 star around the Galactic
Centre are, in principle, a very straightforward tool in order to
test any theory of gravity. In \cite{capo12}, there is an overview
of self-gravitating structures, at different scales, whose dynamics
could be described without asking for dark matter. According to
\cite{capo12}, the relations between $f_1$, $f_2$ and $\delta$ and
$\Lambda$ parameters are $f_1=1+\delta$,
$f_2=-(1+\delta)/(\Lambda^2)$.

Specifically, we have to find the minimal values of the reduced
$\chi^{2}$ in order to determine $f_1$ and $f_2$ assuming $f_0 = 0$.
This allows to reconstruct $f(R)$ models up to the second order.

Fig. \ref{fig01} presents the maps of the reduced $\chi^{2}$ over
the $\{f_1-f_2\}$ parameter space for all simulated orbits of S2
star which give at least the same or better fits to the NTT/VLT
observations of S2 star than the Keplerian orbits ($\chi^{2}=1.89$).
The upper panel corresponds to $f_1$ in the range $[-25, 0]$, and
the lower panel to the range $[0, 25]$, respectively. We can see
that, in a large region  of the parameter space, $\chi^{2}$ of the
orbits in modified potential is less than the value in Newtonian
potential. However, it seems that we cannot constrain both $f_1$ and
$f_2$ using only the observed S2 orbits because these two parameters
are strongly correlated. We can constrain only their ratio
$f_1/f_2$. According to \cite{capo12}, the effective mass is
$m^2=-f_1/(3f_2)$. The solutions are valid if $m^{2} > 0$ i.e. $f_1$
and $f_2$ are assumed to have different signs.

This is a degeneracy problem that have to be removed in order to
obtain reliable results. Such a problem is also found  in fitting
the flat rotation curve of spiral galaxies. As discussed in details
in \cite{stab13}, the $f_1/f_2$ degeneracy can be removed by  using
potentials coming from $f(R,\phi)$ gravity. In this model, two
potentials, $\Psi(x)$ and $\Phi(x)$, result as entries of the metric
in the Newtonian limit. The combination of both potentials gives
rise to the effective potential (\ref{sanders}) that affects the
particle (in our case the S2 star).

The potential includes the gravitation constant measured at infinity
$G_\infty$. The relation between $G_\infty$ and $G$, the
gravitational constant measured in the laboratory, is given by the
above formula (\ref{G}). If we take $\omega (\phi_0)=1/2$ and use
the relation $1/\alpha = 3-2\omega (\phi_0)\phi_0$, we get
$\phi_0=3-1/\alpha$. Combining these relations, we get: $G_\infty =
G/(1+\alpha)$.

Our aim is to determine $f_0$, $f_R$, $f_{RR}$, $f_{\phi}$,
$f_{\phi\phi}$ and $f_{\phi R}$. For the lowest order of the field,
as we said,  one can set $f_0=0$ and $f_{\phi}$=0. We  use also the
further constrains given in \cite{stab13} at lowest order, that is
$f_{\phi R}=1$, $f_{RR}=0$, and $f_R=\phi_0$. This last relation
gives $f_R=\phi_0=3-1/\alpha$.

For $f(R,\phi)$ gravity, one can define a further effective mass
\cite{stab13}, that is  $m_\phi^2=-f_{\phi\phi}/(2\omega(\phi_0))$
and if we take $\omega(\phi_0)=1/2$, we get immediately
$f_{\phi\phi}= - m_\phi^2$.

Finally we can assume the following set of parameters $f_0=0$,
$f_R=3-1/\alpha$, $f_\phi=0$, $f_{RR}=0$, $f_{\phi R}=1$ and
$f_{\phi\phi}=- m_\phi^2$.
These choices are physically reliable and mean that we can assume an asymptotic  Minkowski background, i.e.  $f_0=0$, that the General Relativity is recovered for $f_\phi=0$, $f_{RR}=0$, $f_{\phi R}=1$, and effective massive modes (and then effective lengths) are related to $f_R=3-1/\alpha$,  and
$f_{\phi\phi}=- m_\phi^2$. In particular, $f_0=0$ means that cosmological constant can be discarded at local scales.

Figs. \ref{fig02} and \ref{fig03} are the maps of the reduced
$\chi^{2}$ over the $\{\alpha - m_\phi\}$ parameter space in
$f(R,\phi)$ gravity for all simulated orbits of S2 star which give
at least the same or better fits than the Keplerian orbits
($\chi^{2}=1.89$). The upper panel in Fig. \ref{fig02}
corresponds to $m_\phi$ in $[0, 0.06]$ and $\alpha$ in $[0, 0.33]$,
and the lower panel to the zoomed range of $m_\phi$ in $[0, 0.03]$
and $\alpha$ in $[0, 0.05]$, respectively.
For $\alpha < 0$, there is no region in the parameter
space where $\chi^{2}<1.89$ (Keplerian case). For $0 < \alpha < 1/3$ there are
two regions where $\chi^{2}<1.89$ (for $m_\phi < 0$ and $m_\phi >
0$), but the absolute minimum is for $m_\phi < 0$. We obtained absolute
minimum of the reduced $\chi^{2}$ for $\alpha$ in the interval
$[0.0001, 0.0004]$, and $m_\phi$ in the interval $[-0.0029,
-0.0023]$ (see Fig. \ref{fig03}). The absolute minimum of the
reduced $\chi^{2}$ ($\chi^{2}=1.5011$) is obtained for
$\alpha$ = 0.00018 and $m_\phi$ = -0.0026, respectively.

The simulated orbits of S2 star around the Galactic Centre in Sanders
gravity potential (blue solid line) and in Newtonian gravity
potential (red dashed line) for $\alpha$ = 0.00018
and $m_\phi$ = -0.0026 during 10 periods, are presented in Fig.
\ref{fig04}. We can notice that the precession of S2 star orbit has
the same direction as in General Relativity. The precession of S2
star orbit in the same direction can be also obtained for some
ranges of parameter $\delta$ in general Yukawa potential (for
more details see paper \cite{bork13}). As it can be read from
Fig. \ref{fig04}, the best fit orbit in Sanders gravity potential
precesses for about $3^\circ.1$ per orbital period.

In case of Sanders potential, analytical calculation of orbital
precession is very complicated to obtain, so we calculated it
numerically and presented in Figs. \ref{fig05} and \ref{fig06}
as a function of $\alpha$ and $m_\phi$.
Assuming that a potential does not differ significantly from
Newtonian potential, we derive perturbing potential from:

\begin{equation}
V(r) = \Phi \left( r \right) - {\Phi_N}\left( r \right)\begin{array}{*{20}{c}}
;&{{\Phi_N}\left( r \right) =  - \dfrac{{GM}}{r}}
\end{array}.
\label{equ04}
\end{equation}

\noindent Obtained perturbing potential is of the form:

\begin{equation}
V(r) = - \dfrac{GM \alpha}{r(1 + \alpha)}\left({{e^{-\sqrt{1-3
\alpha} \cdot m_\phi \cdot r} } - 1} \right),
\label{equ05}
\end{equation}

\noindent and it can be used for calculating the precession angle
according to the equation (30) from paper \cite{adki07}:

\begin{equation}
\Delta \theta = \dfrac{-2L}{GM e^2}\int\limits_{-1}^1 {\dfrac{z \cdot dz}{\sqrt{1 - z^2}}\dfrac{dV\left( z \right)}{dz}},
\label{equ06}
\end{equation}

\noindent where $r$ is related to $z$ via: $r = \dfrac{L}{1 + ez}$.
By differentiating the perturbing potential $V(z)$ and substituting
its derivative and expression for the semilatus rectum of the
orbital ellipse ($L = a\left( {1 - {e^2}} \right)$) in above
equation (\ref{equ06}), and taking same values for orbital elements
of S2 star like in paper \cite{bork12} we obtained numerically for
$\alpha$ = 0.00018 and $m_\phi$ = -0.0026 that precession per
orbital period is $3^\circ.053$.

Graphical presentation of precession per orbital period for $\alpha$
in the range $[0.0001, 0.0003]$ and $m_\phi$ in $[-0.003, -0.002]$
is given in Fig. \ref{fig05}, and the case for $\alpha$ in
$[-0.0005, 0.0005]$ and $m_\phi$ in $[-0.003, -0.0025]$ is presented
in Fig. \ref{fig06}. As one can see pericenter advance (like in GR)
is obtained for positive $\alpha$, and retrograde precession for
negative $\alpha$. However, it should be taken into account that
fits better than Keplerian are obtained only for positive $\alpha$
and hence for the precession in the same direction as in GR.

General Relativity predicts that pericenter of S2 star should advance
by $0^\circ.08$ per orbital revolution \cite{gill09b} which is
much smaller than the value of precession per orbital period in
Sanders gravity potential, but the direction of the precession is
the same.

\section{Conclusions}

In this paper, we compared the observed and simulated S2 star orbits
around the Galactic Centre, in order to constrain the parameters of
gravitational potentials derived from modified gravity models. The obtained
results are quite comfortable for the effective gravitational
potential derived from $f(R,\phi)$ gravity that, essentially,
reproduce  Sanders-like potentials \cite{sanders1,sanders2}
phenomenologically adopted to explain the rotation curves of
spiral galaxies. Also if such kind of potentials are not sufficient in  addressing completely  the problem of dark matter in galaxies \cite{dubal}, they give indications that alternative theories of gravity could be viable in describing galactic dynamics.

In other words, orbital solutions derived from such
a potential are in good agreement with the reduced $\chi^2$ deduced
for Keplerian orbits. This fact allows to fix the range of variation
for $\alpha$ and $m_\phi$, the two parameters characterizing the
potential (\ref{sanders}). The precession of S2 star orbit obtained
for the best fit parameter values ($\alpha$ = 0.00018 and $m_\phi$ =
-0.0026) has the positive direction, as in General Relativity.

In particular, we fitted the NTT/VLT astrometric observations of S2 star, which contain
a possible indication for orbital precession around massive compact object
at Galactic centre, in order to constrain the parameters of Sanders-like
gravity potential, since this theory has not been tested at these scales
yet.  We obtained much larger orbital precession of S2 star in Sanders-like
gravity than the corresponding value predicted by General Relativity.
In the paper \cite{gill09b}, the authors presented Newtonian orbit but with moved position
of central point mass in different way. In that way they explained
observed precession. In our calculated orbit by Sanders like potential for
best fitting parameters, we also obtained precession but with fixed
position of central point mass, since we have not to move it.

However, one should keep in mind that we considered an idealized
model ignoring many uncertainty factors, such as  extended mass
distributions, perturbations from non-symmetric mass distribution,
etc. Therefore, future observations with advanced facilities, such
as GRAVITY which will enable extremely accurate measurements of the
positions of stars of $\sim$ 10 $\mu$as \cite{gill10a}, or E-ELT
with expected accuracy of $\sim$ 50-100 $\mu$as \cite{eelt}, are
needed in order to verify the claims in this paper. As final remark,
we believe that the surveys aimed to give details on the dynamics
around the Galactic Centre could be a powerful tool to test theories
of gravity.


\begin{acknowledgments}
S.C. acknowledges the support of INFN ({\it iniziativa specifica} TEONGRAV).
D.B., P.J. and V.B.J. wish to acknowledge the support by the
Ministry of Education, Science and Technological Development of the
Republic of Serbia through the project 176003.
\end{acknowledgments}


\end{document}